\documentclass[preprint,5p]{elsarticle}

\usepackage{units,amsmath,amssymb,bm,soul}
\usepackage{subfigure}
\usepackage{epsfig}

\usepackage{graphicx,xcolor}
\usepackage{epstopdf}

\usepackage{eurosym}  

\usepackage{float}

\newcommand{\be}{\begin{equation}}
	\newcommand{\ee}{\end{equation}}

\newcommand{\bea}{\begin{eqnarray}}
	\newcommand{\eea}{\end{eqnarray}}

\newcommand{\half}{\dfrac{1}{2}}

\journal{Combustion and Flame}

\begin{document}
	
	\begin{frontmatter}

\title{A thermomechanical explanation for the topology of crack patterns observed on the	surface of charred wood and particle fibreboard}

\author[aalto]{Djebar Baroudi}

\author[aalto,tallinn]{Andrea Ferrantelli}%

 \author[aalto]{Kai Yuan Li}
\cortext[cor1]{Corresponding author}
\ead{kaiyuan.li@aalto.fi}

 \author[aalto]{Simo Hostikka}

\address[aalto]{Aalto University, Department of Civil Engineering,	P.O.Box 12100, 00076 Aalto, Finland	}

\address[tallinn]{Tallinn University of Technology, Faculty of Civil Engineering, 19086 Tallinn, Estonia}

\begin{abstract}
	
	In the assessment of wood charring, it was believed for a long time that physicochemical processes were responsible for the creation of cracking patterns on the charring wood surface. This implied no possibility to rigorously explain the crack topology. In this paper we show instead that below the pyrolysis temperatures, a primary global macro-crack pattern is already completely established by means of a thermomechanical instability phenomenon. First we report experimental observations of the crack patterns on orthotropic (wood) and isotropic (Medium Density Fibreboard) materials in inert atmosphere. Then we solve the 3D thermomechanical buckling problem numerically by using the Finite Element Method, and show that the different crack topologies can be explained qualitatively by the simultaneous thermal expansion and softening, taking into account the directional dependence of the elastic properties. Finally, we formulate a 2D model for a soft layer bonded to an elastic substrate, and find an equation predicting the inter-crack distance in the main crack-pattern for the orthotropic case. We also derive a formula for the critical thermal stress above which the plane surface will wrinkle and buckle. The results can be used for finding new ways to prevent or delay the crack formation, leading to improved fire safety of wood-based products.
\end{abstract}

\date{\today}

\end{frontmatter}

\section{Introduction}

Burning wood and other cellulosic materials leave behind a char layer that usually shows a distinct pattern of cracks.
During combustion, the char acts as a heat barrier between the combustion zone and virgin material, reducing the burning rate of the uncharred material \cite{DiBlasi200847}. The shrinkage and cracking of the char layer have been found to influence the pyrolysis \cite{Bryden20031633,FAM2207} and the fire resistance of structures made of cellulosic materials. For instance, the char layer reduces the heat transfer to the virgin material while the cracks enhance it due to the flame attachment to the crack locations \cite{Li201539}. Roberts \cite{ROBERTS197179} reported that the charring process of wood took place at approximately 370$^\circ{\mathrm C}$, and that the charred residues greatly influence the combustion processes because they form barriers between flames and unburnt material. However, the charred surfaces exhibit patterns of cracks that will weaken this barrier function. Both mechanisms of crack formation and its effects on combustion process are still unknown. According to Babrauskas \cite{Babrauskas2005528}, still a few decades ago the arson investigators used to attribute the type and spacing of the cracks to the heating rate and possible use of liquid accelerants, but the belief was disproved by Ettling \cite{Ettling} who showed, using furnace experiments, that the exposure temperature alone could explain the observed pattern.

\begin{figure}[t]
	\centering
	\includegraphics[width=0.45\textwidth]{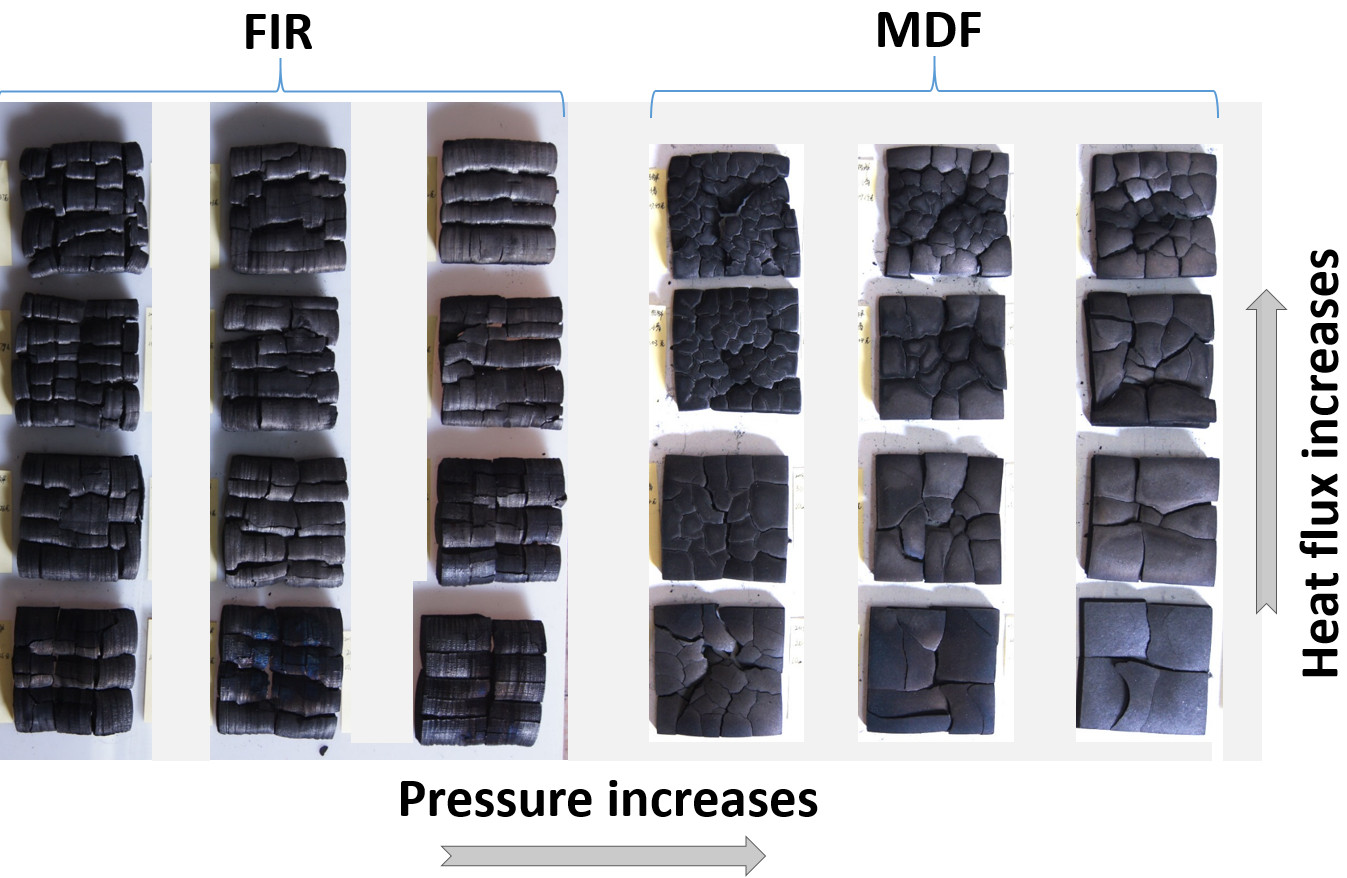}
	\caption{Crack patterns at the end of the experiment. 
		Left: plain wood (orthotropy), right: fibreboard (isotropy).}
	\label{fig:physical_problem_patterns}
\end{figure}

Attempts have been made to model (or explain) the crack-patterns by drying processes, i.e. where the material shrinkage is the driving force \cite{Ross}; however such theories cannot explain the characteristic crack patterns observed on charred surfaces of wood and fibreboard (Fig.\ref{fig:physical_problem_patterns}). In the studies of wood pyrolysis, i.e. the thermal decomposition of solid into flammable liquids and gases and solid char, attention has mainly focused on physicochemical processes taking place during and beyond pyrolysis, ignoring the thermomechanical processes occurring below the pyrolysis temperature $T_p\approx 300 ^\circ{\mathrm C}$ \cite{FAM2207,Li201539,doi:10.1080/10618562.2012.659663,Stoliarov20102024,Lautenberger20091503}. Understanding the mechanisms of char cracking can thus lead to improved accuracy of the combustion modelling of wooden materials, enabling more profound considerations of charring behaviours in fire safety and energy technologies.
Our observations of cracking phenomena (Section~\ref{sec:experimental-analysis}) show that i) the entire main crack-pattern forms suddenly at all locations and at the same time on the surface, ii) this occurs \textit{before} any actual charring and iii) the patterns are quasi-periodic. These are typical features of a mechanical instability phenomenon of buckling \cite{bloom2000handbook,Cirak1932,Perzyna1994,Benallal2004725,Bogdanovich2009}.

At temperatures close to but below the onset of pyrolysis at $T_p$, wood is indeed a natural thermoplastic material \cite{Salmen,Antoniow2012}. The glass transition temperature of dry wood has been observed to be around $T_g\approx 200 ^\circ{\mathrm C}$, at which it simultaneously softens and elongates extensively. Restrain thermal stresses are then induced in the hot layer, and under certain conditions that we explain in Section \ref{sec:approximate-2d-solution-over-a-thin-plate}, this eventually leads to wrinkling \cite{Salmen,Bazant_1985,Salmen_1984,Antoniow2012}.

In this work, we want to find out if the cracking patterns on charred surfaces can be induced by a thermomechanical buckling of a hot thin soft surface layer bonded to a cold, harder elastic substrate (elastic foundation). To investigate the problem, we formulate a model for the experimental conditions where char cracking was observed, and solve it both analytically and numerically to compare the resulting buckling modes with the experimental observations of crack patterns. We do not investigate the crack initiation and propagation processes themselves but focus on the underlying instability phenomenon, to explain the instantaneous and periodic crack patterns. Damage and fracture mechanics studies may be needed in the future, when the effectiveness of the possible means to reduce or delay the crack formation will be evaluated.

\begin{figure}[h]
	\centering
	\includegraphics[width=0.4\textwidth]{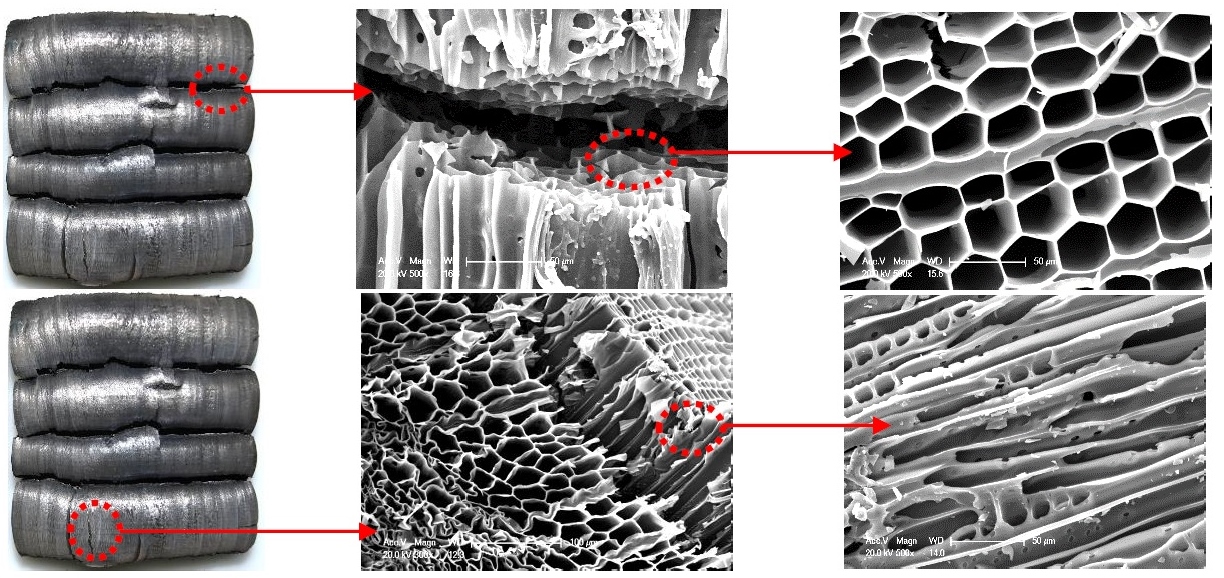}
	\caption{Main crack patterns perpendicular to the fibres \cite{Li2016}.}
	\label{fig:firfissures}
\end{figure}

As we will show, our thermomechanical model not only reproduces the primary patterns for fir and fibreboard in Fig.\ref{fig:physical_problem_patterns}. It also explains why the cracks observed for fir (Fig.\ref{fig:firfissures}) are mainly perpendicular to the fibres, where the mechanical resistance properties are stronger, and not parallel, as one would expect by considering only wood shrinkage.

The present paper is organized as follows: in Section \ref{sec:experimental-analysis} we discuss the experimental setup and measurements, while in Sections \ref{sec:full-3d-non-linear-thermomechanical-model} and \ref{sec:results-of-3d-simulation} we construct a non-linear 3D model that we solve numerically. In Section \ref{sec:approximate-2d-solution-over-a-thin-plate} we perform a dimensional reduction from 3D to 2D to formulate a model for the buckling of a thin layer bonded to an elastic substrate, and derive analytical formulas to identify the controlling parameters of the buckling process. Section \ref{sec:conclusions} is dedicated to our conclusions and \ref{sec:analytical-solution-for-buckling-modes-of-a-thin-wood-layer}
 contains the details of the analytical 2D model, both for the torsion-free and the coupled case.


\section{Experimental analysis}\label{sec:experimental-analysis}

\begin{center}
	\begin{figure}[t]
		\centering
		\includegraphics[width=0.45\textwidth]{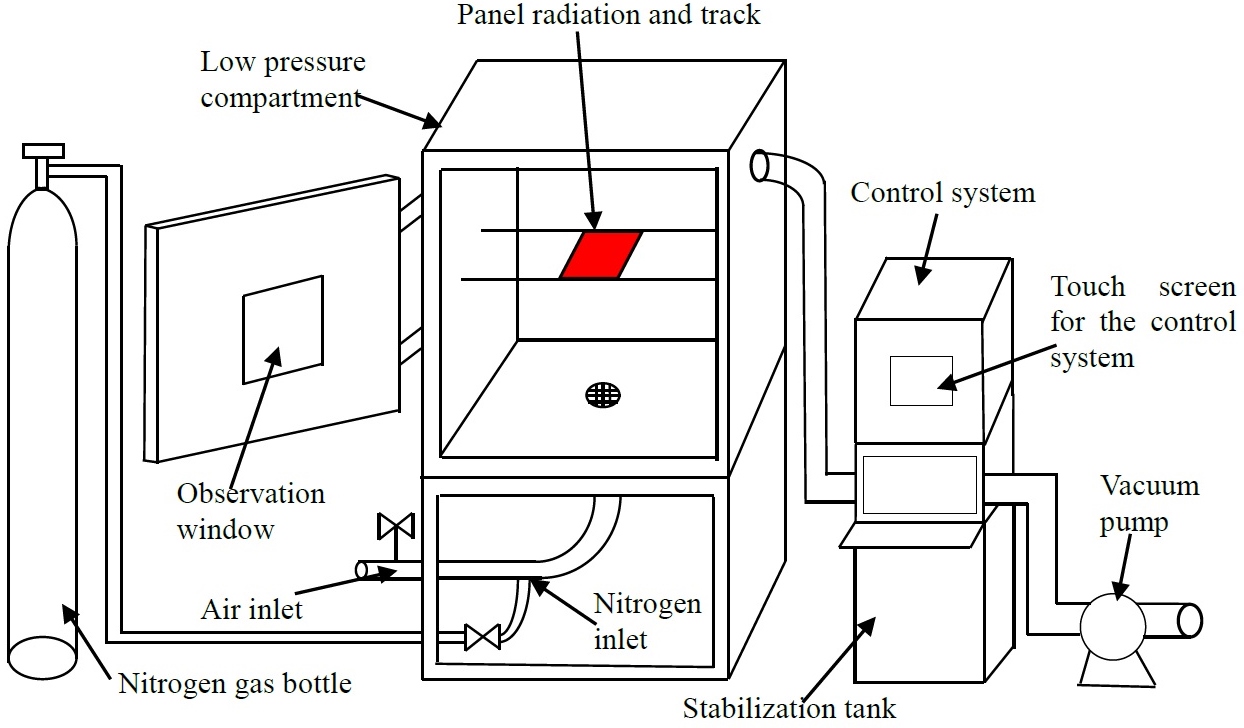}   
		\caption{Sketch of experimental rig \cite{Li2016}.}
		\label{fig:Experimentalrig}
	\end{figure}
\end{center}

\subsection{Experimental setup}\label{Experimental setup}

\begin{center}
	\begin{figure*}[h!]
		\centering
		\includegraphics[width=0.70\textwidth]{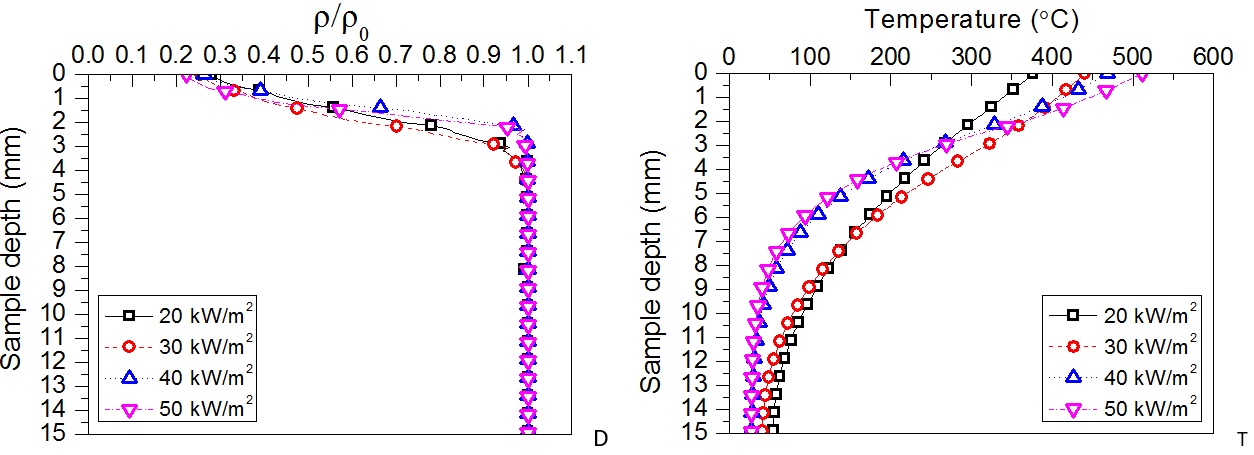}
		\caption{Instantaneous profiles of relative density (left) and temperature (right) at the moments of crack formation.}
		\label{fig:Temperatureprofile}
	\end{figure*}
\end{center}

Pyrolysis experiments were carried out to measure the surface cracking and charring behaviours of planar samples of construction materials, by exposing them to external heat fluxes simulating fire conditions. Nitrogen atmosphere was used to prevent surface oxidation reactions that would transform the charred surface into ash. The experimental rig (Figure \ref{fig:Experimentalrig}) consisted of three major components: the gas supply system, a low pressure compartment and the control system. The gas supply system provided nitrogen stored in a bottle and air from ambient environment through two pipes. A valve on each pipe controlled the gas flow rate.

The low pressure compartment was 1m long, 1m high and 0.6m wide. Inside the compartment, a 0.3m x 0.3m panel radiator was used to generate uniform radiative heat fluxes towards the samples. The panel radiator was an infrared panel heater that could generate an incident heat flux up to 100 kW/m$^2$, with the highest temperature being 1800$^o$C. More details are available in \cite{Li2016}.

The panel radiator was hanged at two tracks to adjust its position. The vertical distance between the panel and sample surfaces was 30mm. The sample holder was made of Kaowool, and an electric balance was used to weight the sample while two thermocouples measured the temperatures both inside the sample and on its surface. A digital camera was placed in front of the observation window to record the charring process. The control system included a stabilization tank, a vacuum pump and a central control unit with a touch screen to manage the experimental process.

The charred samples were sent for SEM experiments to examine the microstructures of char and char fissures after pyrolysis. The SEM equipment is XL30 ESEM-TMP made by Royal Philips of the Netherlands, it can be used for micro-morphology research of electrically conductive solid materials.

\subsection{Materials}\label{Materials}

The current study uses Medium Density Fibreboard (MDF), made by a local manufacturer, and natural fir.
The MDF samples are 100mm long, 100mm wide, 15mm thick with a bulk density $\rho_{\rm MDF}=730\pm17$ kg/m$^3$. The fir samples are 100mm long, 100mm wide, 25mm thick with a bulk density $\rho_{\rm fir}=363\pm18$ kg/m$^3$. According to the manufacturer, the MDF panels are made from pine tree, with approximately 10\% of resin and wax as the additives. To evaluate the grain effect of natural wood, the fir samples were cut parallel to the grains and symmetrically with respect to the centre of the annual rings.

The samples were dehydrated in an oven at 95-100$^o$C for at least 24 hours to remove the moisture. Their mass was monitored every 2 hours during the drying process to ensure mass stabilization. The samples were then sealed in plastic bags and weighted again after cooling, to ensure that any change in the moisture content was insignificant. After the pyrolysis experiments, the charred samples were split to small particles with or without fissures for SEM experiments. As the charred samples were not electrically conductive, they were treated by splashing conductive material at the surfaces before the SEM experiments.

\subsection{Experimental conditions and procedure}\label{sec:experimental-conditions-and-procedure}

Four incident heat fluxes were used in the pyrolysis experiments: 20, 30, 40 and 50 kW/m$^2$,
while three ambient pressures of 30, 60 and 95 kPa were applied. Before the experiments, the uniformity of the heat flux distribution was evaluated by measuring the incident heat flux at different points at the same level of the sample surface.
We found that within 100mm of distance, the heat fluxes deviate less than 2.7\% from the nominal value.

During the experiments, a vacuum pump first reduced the absolute pressure in the compartment to 5 kPa to remove most of the air, then pure nitrogen was led in from the bottom of the compartment at a flow rate of 0.6 m$^3$/min. The vacuum pump was off until the internal pressure reached the target experimental pressure, and then turned on again to stabilize it.

A voltage control was turned on simultaneously with the vacuum pump, to adjust the temperature of the panel radiator. Once the temperature got stabilized at its target value for 5 minutes, the radiator was moved by the control system along the track to the position right above the sample centre. To protect the compartment from deformation, no pressure lower than 30 kPa was used, apart from the instantaneous low pressure for air removal.

\subsection{Experimental observations}\label{sec:experimental-observations}

Two distinct macro-crack pattern types are identified on the sample surface (Fig.\ref{fig:physical_problem_patterns}): 1) For fir, the cracks are formed as a 1D pattern perpendicular to the wood grain direction. The distance between the cracks is roughly constant in each of the experiments. 2) The MDF, in turn, shows 2D-patterns with mirror symmetries across the diagonals of the square samples. 

The cracking times were determined by eye inspection of the sample surface. The observed times are presented in \cite{Li2016}. A close examination of the video material revealed that the large cracks appeared on the sample surface simultaneously.

\begin{figure*}[t]
	\centering
	\includegraphics[width=\textwidth]{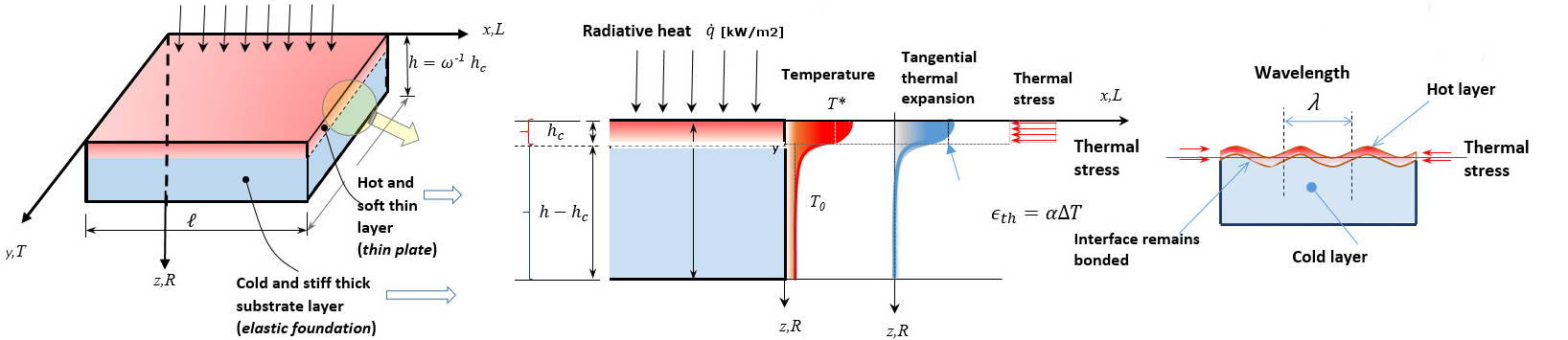}	
	\caption{A schematic of the (hot) thin wood plate and the thick (cold) elastic substrate: temperature definitions and tangential thermal expansion profile together with the bi-axial thermal stresses.
	}
	\label{fig:Thinplate}
\end{figure*}

\subsection{Thermal analysis}\label{sec:thermal-analysis}

The pyrolysis model in Fire Dynamics Simulator (FDS) version 6.3.2 \cite{mcgrattan2013fire} is used for calculating the temperature and density profiles inside the samples during the experiments described above. The model solves the coupled heat conduction and pyrolysis reaction equations using a one-dimensional finite difference method.  Assuming a single-step conversion reaction for pyrolysis, the governing equations for the densities of the virgin material $\rho$ and char $\rho_c$, the heat conduction equation with source term $\dot{q}_s^{'''}$, and the heat production (loss) rate $H_r$ are \cite{mcgrattan2013fire}

\bea
&&
\frac{d \rho}{dt}= -\rho_0 r\,, \\
\label{massConservation}
&&
\frac{d \rho_c}{dt} = \nu_c \rho_0 r \,,
\\
&&
\rho c\dfrac{\partial T(z,t))}{\partial t}=\dfrac{\partial }{\partial z}\left(k \dfrac{\partial T(z,t)}{\partial t}\right)+\dot{q}_s^{'''}\,,
\label{heatconduction}
\\
&&
\dot{q}_s^{'''}=-\rho_{0} r H_{r}\,,
\label{sourceterm}
\eea
where $T(z,t)$ is the solid temperature at depth $z$ from the surface, $c$ and $k$ are respectively the specific heat capacity and thermal conductivity, and $\nu_c$ is the char yield of the conversion reaction. $\rho_0$ is the initial density of the virgin material. The reaction rate is calculated as
\bea
&&
r=\left(\dfrac{\rho}{\rho_{0}} \right)^{n_{s}}
A_{s}
\exp\left(-\dfrac{E_{s}}{RT} \right)\,,
\label{reactionrate}
\eea
with a kinetic parameter $A_s\,[s^{-1}]$, $n_{s}$ the reaction order and $E_{s}$ the activation energy.

Assuming appropriate initial and boundary conditions as well as thermal and kinetic parameters \cite{mcgrattan2013fire,doi:10.1080/10618562.2012.659663}, we can solve the temperature and density profiles at different times, as shown in Fig.~\ref{fig:Temperatureprofile} for four different heat fluxes. 
As the crack formation times were associated with large uncertainty, the profiles are plotted at post-cracking times when the cracks were certainly created and visible by eye. The layer at which $T > T_g$ is relatively thin in comparison to the sample thickness. The very surface of the material has already reached the pyrolysis temperature.

For the purpose of the mechanical modelling, the induced temperature distribution along the wood thickness $z$ can be approximated by two zones: one hot and thin above, and the other colder and thicker below.


\section{Full 3D non-linear thermomechanical model}
\label{sec:full-3d-non-linear-thermomechanical-model}

The physical problem of surface wrinkling can be investigated through a 3D thermoelasticity formulation. Considering only the effect of thermal stresses, this leads to an eigenvalue problem for a system of three-dimensional thermoelasticity equations on a thin elastic layer. The in-plane thermal elongations of this layer are restrained by an elastic substrate to which it is perfectly bonded (Fig.~\ref{fig:Thinplate}). The three-dimensional thermoelasticity equilibrium equations are
\be\label{equilibriumeq}
{\rm div} {\boldsymbol\sigma} + \rho {\bf f} = {\bf 0}\,,
\ee
where ${\boldsymbol\sigma}$ is the stress tensor of the material, $\rho$ its density and ${\bf f}$ the resultant of pressure forces.

The stress tensor of the material is written according to the general Hook's law as
\be\label{hook}
\boldsymbol\sigma =  {\bf D}: \left(\boldsymbol\epsilon - \boldsymbol{\epsilon}^{(\rm th)}\right) \,,
\ee
where ${\bf D}$ is the symmetric elasticity tensor and $\boldsymbol{\epsilon}^{(\rm th)}$ is the thermal strain tensor. The total deformation or strain tensor $\boldsymbol\epsilon$ can be explicitly written in terms of the displacement ${\bf u}$,
\be\label{straindisplacement}
\boldsymbol\epsilon = {1 \over 2} [(\nabla {\bf u})^{\rm T}  + \nabla {\bf u} +   (\nabla {\bf u})^{\rm T} \nabla {\bf u} ]\,,
\ee
and also decomposed into thermal and elastic strains $\boldsymbol\epsilon^{(\rm e)}$,
\be\label{totaldeformation}
\boldsymbol\epsilon = \boldsymbol\epsilon^{(\rm th)} + \boldsymbol\epsilon^{(\rm e)}
=
\boldsymbol\alpha  \Delta T + \boldsymbol\epsilon^{(\rm e)}\,,
\ee
where we recognise the thermal expansion tensor $\boldsymbol\alpha$, which is diagonal, and the temperature change $\Delta T$.

The main unknown is the thermally induced displacement field ${\bf u}$ with components $(u,v,w)$ along the axes $(x,y,z)$. The membrane state (pre-critical state) corresponds only to in-plane motion ($w\equiv 0$), while in buckling conditions, bending produces off-plane motion $w \ne 0$. To solve the Cauchy problem, appropriate boundary conditions are assumed.

In our application, we consider only the action of thermal stresses in the equilibrium equation (\ref{equilibriumeq}), namely we set the external forces to zero, $\bf f = 0$. This leads to an eigenvalue problem. In particular, the nodes of the eigenmodes coincide with
the locations of the most stressed (or equivalently, stretched) loci on the surface of the thin layer\footnote{As we will show in Section \ref{sec:approximate-2d-solution-over-a-thin-plate}, at such locations the major extensional stress is maximal.}. These are the locations where cracks should initiate when the mechanical resistance of the material decreases with increasing temperature.


\section{Results of 3D simulations}\label{sec:results-of-3d-simulation}

An exact analytical solution of the full 3D eigenvalue problem given by Eqs.(\ref{equilibriumeq}), (\ref{hook}), (\ref{straindisplacement}) and (\ref{totaldeformation}) cannot be obtained, though some approximations can be made (for instance, by means of the Rayleigh-Ritz method \cite{Leissa2005961,Trefethen:1997:NLA,trove.nla.gov.au/work/6379829}). 
We therefore solve it numerically for fir and MDF by Finite Element Method (FEM), with the program COMSOL Multiphysics \cite{Hanke06}.

To define the problem, consider Fig.~\ref{fig:Thinplate}, where the longitudinal $L$, transverse $T$ and radial $R$ directions of the sample correspond respectively to the axes $(x,y,z)$. During heating, the surface temperature rises from $T_0$ to $T^*$. The FEM solutions are obtained in two stages: first calculating the membrane stress state resulting from thermal expansion, then running a separate buckling analysis with different values of the mechanical properties.

An orthotropic elastic model is used for fir, while for MDF we use an isotropic elastic model. The fir mechanical properties were taken from \cite{Woodhandbook}, with the Poisson's ratios $\nu_R=0.08$, $\nu_T=0.5$ and $\nu_L=0.4$, the Young moduli $E_R(T_0) = 0.1$ GPa, $E_R(T^*)=0.05$ GPa, $E_T(T^*)=0.03$ GPa and $E_L(T^*)=0.5$ GPa. These give the characteristic ratios $E_R(T_0) / E_L (T^*) = 0.2$ and $E_R(T_0) / E_T (T^*) = 3.33$. For a typical wood, $E_L(T)/E_T(T)\approx 0.1$ and the thermal expansion coefficients along the three axes hold as follows,
\bea
\alpha_R&=&(12\div17)\times 10^{-6}\dfrac{1}{K}\,,\label{alphaR}\\
\alpha_L&=&(1.6\div2.4)\times 10^{-6}\dfrac{1}{K}\,,\label{alphaL}\\
\alpha_T&=&(16\div24)\times 10^{-6}\dfrac{1}{K}\label{alphaT}\,,
\eea
namely $\alpha_R\approx \alpha_T= 10\times\alpha_L$ \cite{Woodhandbook}.
\begin{figure}[t]
	\centering
	\includegraphics[width=0.48\textwidth]{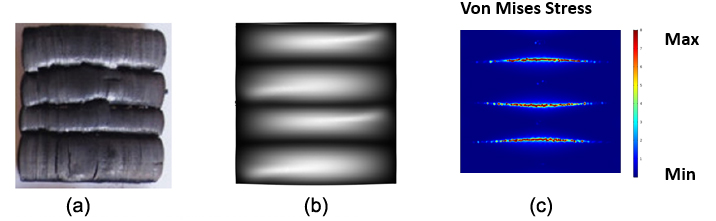}
	\caption{Experimental (a) and numerical 3D results for fir surface wrinkling as off-plane displacement (b) and Von Mises stress (c).}
	\label{fig:orthotropicvonmises}
\end{figure}

 Figure~\ref{fig:orthotropicvonmises} shows a comparison of experimental crack patterns for fir (a), the numerical solutions in terms of the absolute value of off-plane displacement (b) and the von Mises stress (c). In Fig.~\ref{fig:orthotropicvonmises}b, the white colour indicates high displacement values, while zero displacements are in black. The nodes of the buckling mode correspond to the locations of the zero displacement. We can observe that the cracks are orthogonal to the main fibre direction $L$ in both experimental and numerical results. In addition, the experimental crack locations (a) coincide with the computed locations of the maximal stress (c), that in turn coincide with the nodes of the buckling modes (b). 

\begin{figure}[h]
	\centering
	
	\includegraphics[width=0.25\textwidth]{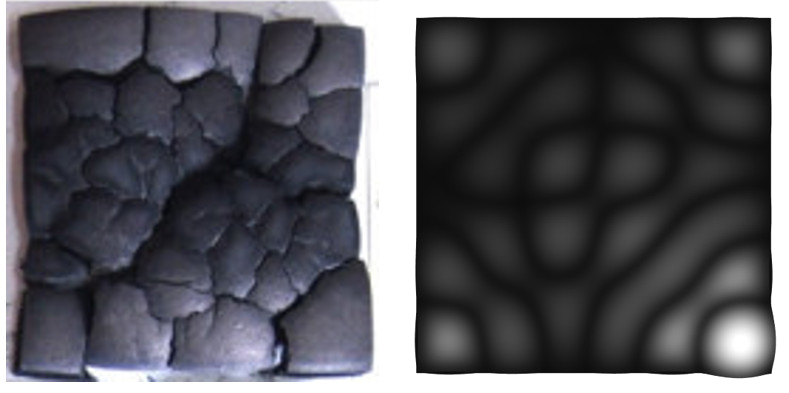}
	
	\caption{Experimental crack pattern (left) and numerical surface wrinkling for isotropic material (MDF).
	}
	\label{fig:Model_versus_TEST_ISO_simple}
	
\end{figure}

A similar comparison for the isotropic case (MDF), where $E_L(T) =E_T(T) \equiv E(T)$, is shown in Fig.~\ref{fig:Model_versus_TEST_ISO_simple}. The numerical results were calculated assuming 
$E(T_0) / E (T^*) = 3$. Unlike the orthotropic case, the crack patterns are now much more complex and smeared into two directions. However, the crack locations seem to coincide with the nodes of the buckling mode, as they did for natural wood.
\vspace{-1mm}
%
%

\begin{figure}[h]
	\centering
	\includegraphics[width=0.4\textwidth]{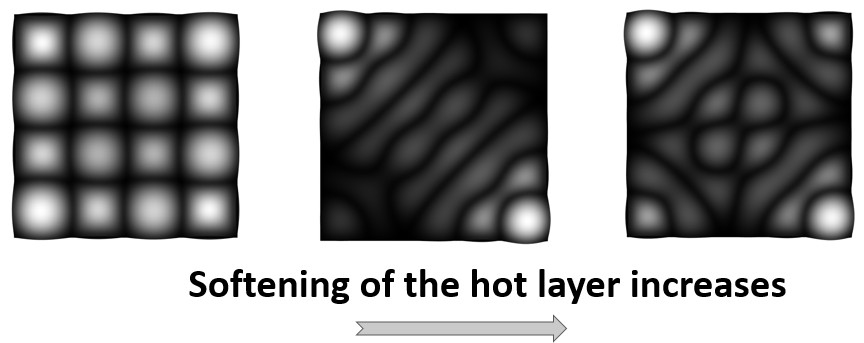}
	\caption{Effect of the increased softening to the observed buckling mode in the numerical 3D results for isotropic case (MDF).
	}
	\label{fig:Transition}
\end{figure}

The sensitivity of the isotropic buckling mode to the relative softening of the material was investigated
by varying the ratio $E(T_0)/E(T^*)$ from 1.0 (no temperature-induced softening) to 2 and 3. The results are shown in Fig.~\ref{fig:Transition}, indicating a transition from a chess-board mode to a diagonally dominated symmetry. In addition to the changing symmetry, local modes of decreasing width seem to appear towards the centre of the sample.
As the mechanical properties of the materials are not exactly known, the precise reproduction of the experimental patterns cannot be expected. However, the qualitative agreement between various experimental patterns and the numerical wrinkling modes with $E_R(T_0)/E_i(T^*) > 1$ for both orthotropic and isotropic materials indicates that the thermomechanical model has captured a major driving mechanism.


\section{Approximate 2D solution over a thin plate}\label{sec:approximate-2d-solution-over-a-thin-plate}

The buckling problem of a thin plate bonded to an elastic substrate allows harmonic solutions that are similar to our experimental observations \cite{Cai20111094,Li20122077,Audoly20082444,hutchinson1992mixed,Bowden1998,B802848F,B814145B,C2SM06844C}. The thin-plate approximation is valid when the thickness of the hot layer is $\lesssim 0.1$ the characteristic width of the sample \cite{ventsel2001thin}, a condition that is satisfied in Fig.~\ref{fig:Temperatureprofile}. The dimensional reduction into 2D problem can be accomplished by the Kirchhoff-Love theory of plates (classical plate theory) \cite{ventsel2001thin}, with the addition of the kinematic connection to the half elastic space (the cold layer). Here we discuss only the main features and results of the model, giving more details in \ref{sec:analytical-solution-for-buckling-modes-of-a-thin-wood-layer}.

Considering the system drawn in Fig.~\ref{fig:Thinplate}, we first derive an explicit formula for the number of half waves $n_i$ at buckling. Recalling Eq.(\ref{totaldeformation}), the total free deformation along the direction $i=L,T$ holds as
\be\label{freedeformation}
\epsilon_i = \epsilon_{th}+ \epsilon_s=\alpha_i \Delta T - \alpha_s \Delta \rho\,,
\ee
with $\alpha_i$ given by Eqs.(\ref{alphaR})-(\ref{alphaT}).
Here we can ignore the shrinkage deformation $\epsilon_s =  \alpha_s \Delta \rho\approx0$, since we consider temperatures below pyrolysis and our samples are oven dry. Figure \ref{fig:Temperatureprofile} shows that before pyrolysis, the density variation in the samples is indeed negligible. Restraining the deformations in (\ref{freedeformation}) induces the membrane thermal stresses $N_{xx}$ if $i=L$ and $N_{yy}$ if $i=T$. The buckling occurs as a transition from the pre-critical state of a straight perfect plane (membrane bi-axial stress state) to the post-critical state of a wrinkled shape (membrane and bending state). Such transition is governed by a so-called bifurcation point.

The deformations of the thin layer are induced by free thermal expansion, and correspond to this critical point,
\be\label{deformations}
\epsilon_{i,cr} \equiv \alpha_i \Delta T_{cr}  =  r_c^2 \left(\dfrac{n_i}{ \pi}\right)^2 + \beta_k^4 \left(\dfrac{\ell_i} {n_i \pi}\right)^2\,,
\ee
where
$\beta_k = k / {E_L I_c} $, $k$ is the spring coefficient of an idealized elastic substrate (more on this later) and $r_c^2 = I_c / A_c$, with area $A_c$ and momentum of inertia $I_c$. $\ell_i$ is the plate length in the $i=L,T$ direction. The above formula allows to compute the critical values for the driving parameter $\Delta T_{cr}$, alone or in combination with $h_c$. 

Now, in order to find the number of half waves $n_i$, we first need to write down the buckling equation. Let us invoke the Trefftz stability criterion of neutral equilibrium \cite{ventsel2001thin}, which we apply to a thin plate bonded to an elastic substrate (the Kirchhoff-Love approximation), as already discussed. The change in total potential energy of the system described in Fig.~\ref{fig:Thinplate} is written as
%
\be\label{totalenergy}
\Delta {\bm \Pi (\bf u)}=\Delta \mathbb{U}-\Delta \mathbb{W},
\ee
where $\mathbb{U}$ is the strain or elastic energy of the thin plate and its foundation (elastic springs),
$\mathbb{W}$ is the work of external forces, e.g. of hydrostatic pressure $p$ from above, and
${\rm {\bf u}}=(u,v,w)$ is the displacement field.  
We  assume the plate to be a perfectly bonded elastic Winkler foundation modelled by a spring distribution (the cold layer) \cite{trove.nla.gov.au/work/6379829}, thus the displacement ${\bf u}$ is continuous between the springs and the plate.

The elastic energy consists of membrane and bending energies of the plate, together with the elastic energy of an idealized elastic substrate. The spring coefficient $k\,[N/m^3]$ is computed by integrating the Boussinesq's solution for this problem over a $B\times B$ unit square, to account for a realistic mechanical response of the substrate. This holds as \cite{Worku}
\be\label{springcoeff}
k = 2.25 {G  \over {(1-\nu) B} } = \frac{2.25}{2} {E_{zz}  \over {(1-\nu^2) B} } \approx \frac{9}{8}\frac{E_{zz}}{B}\,,
\ee
where $G$ is the shear modulus and the estimate holds since the Poisson's ratio $\nu\sim (0.02\div0.5)$ for most of hard and soft woods \cite{Woodhandbook}. 
$E_{zz}$ (or, alternatively, $E_R$) is the Young modulus in the radial direction.
The Trefftz criterion
\be
\delta{ (\Delta {\bm \Pi} (\bm u)) } = 0\,,
\ee
gives the buckling equation \cite{ventsel2001thin}
\bea
\label{coupledproblem}
&&
D_{x}w_{xxxx}+2H w_{xxyy}+D_{y}w_{yyyy}
\nonumber\\
 && 
 +N_{xx}^0w_{xx}+N_{yy}^0w_{yy}
	+ kw(x,y) =p(x,y)\,,
\eea
where $w_{xx}\equiv\partial^2 w/\partial x^2$ and so on are the derivatives of the vertical displacement $w$, and $N_{xx}^0$ and $N_{yy}^0$ are the membrane thermal stresses immediately before buckling.\footnote{Variation of $\Delta \bm{\Pi}$ with respect to $u$ and $v$ separately shows that the membrane stresses are constant.} $H \equiv (D_{xy}+2D_S)$ is the effective torsional rigidity, with $D_S = G h_c^3/12$ \cite{Woodhandbook}. The bending rigidities in the $x$ and $y$ directions hold as
\bea
D_{x} &=& \dfrac{E_{xx} }{ (1-\nu_x \nu_y)} \left(\dfrac{h_c^3}{12}\right) \approx E_{xx}  \dfrac{h_c^3}{12},\\
D_{y} &=& \dfrac{E_{yy} } { (1-\nu_x \nu_y)} \left(\dfrac{h_c^3}{12}\right) \approx E_{yy}  \dfrac{h_c^3}{12},\\
D_{xy} &=&  \nu_y D_x \approx \nu_y E_{xx}  \dfrac{h_c^3}{12},
\eea
where $\nu_{x,y}$ is the Poisson's ratio and $h_c$ is the thickness of the thin hot layer. Since we only consider the case without external pressure load, we can set $p(x,y)=0$.

Eq.(\ref{coupledproblem}) is the most general differential equation for this mechanism, for both orthotropic (wood, fir) and isotropic case (MDF). The buckling critical membrane forces $N_{\alpha, cr}$ are determined from the smallest eigenvalue of this equation, as we show in \ref{sec:analytical-solution-for-buckling-modes-of-a-thin-wood-layer}.

\begin{figure}[t]
	\centering
	\includegraphics[width=0.48\textwidth]{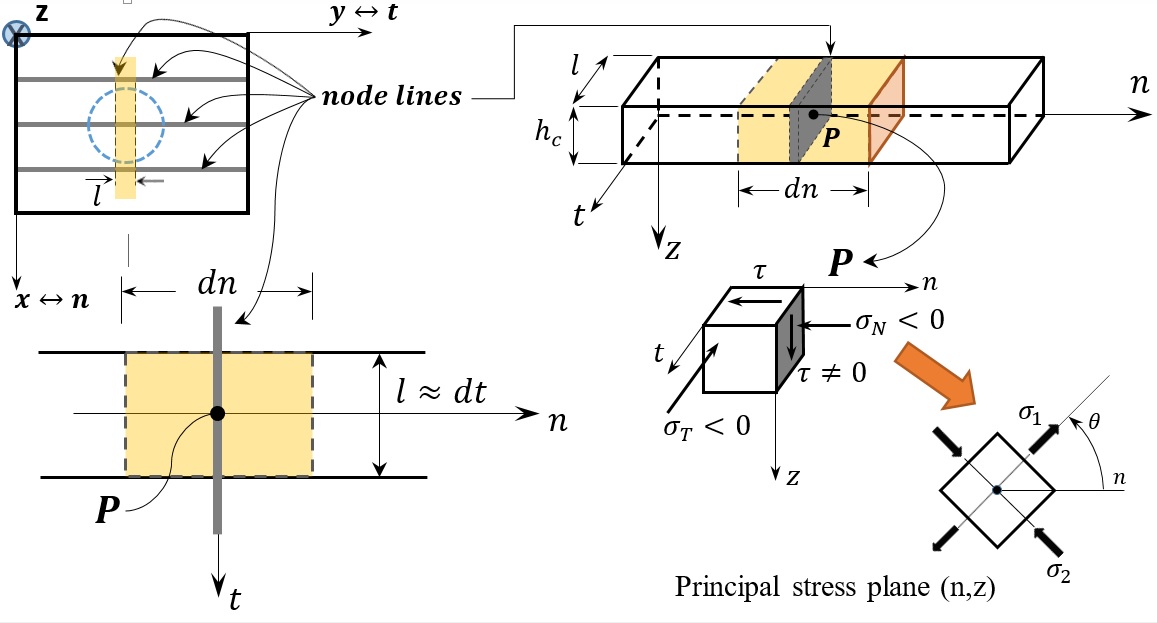}
	\caption{Stress state at a node-point 
		in a section of an infinitely narrow beam orthogonal to the node-line.}
	\label{fig:Extensional_stress}
\end{figure}

Next we focus on the orthotropic case and compute directly the number of half waves $n_i$, which appears in the expression for the deformations Eq.(\ref{deformations}). As the wood's relatively small torsional rigidity $H$ leads to a weak torsion coupling in (\ref{coupledproblem}), we can 
consider the $L$ and $T$ directions separately:
\bea
&&(D_{xx}w^{\prime\prime})^{\prime\prime}+N_{xx}^0w^{\prime\prime}+kw=0\,,\label{decoupled_bucklingL}
\\
&&(D_{yy}w^{\prime\prime})^{\prime\prime}+N_{yy}^0w^{\prime\prime}+kw=0\,,\label{decoupled_bucklingT}
\eea
where $N_{xx}^0$ and $N_{yy}^0$ are the membrane thermal stresses from the pre-buckled state.
Physically, Eqs.(\ref {decoupled_bucklingL}) and (\ref {decoupled_bucklingT}) describe the buckling of plate strips of unit-width $\ell$ on an elastic substrate along the $L$ and $T$ directions. As explained in the Appendix, solving the above differential equations gives the number of half waves for the orthotropic case
\be\label{nalpha}
n_i(T) = 
\dfrac{\sqrt[4]{12}}{\pi}  \dfrac{\ell_i}{ h}   \sqrt[4]{E_R(T_0) 
	\over 
	E_i(T^*)} \sqrt[4]{  { 1\over  { \omega^3 }} {h \over 1{\mathrm m}} }\,,
\qquad
(i=L,T)
\ee
where $h$ is the sample thickness, written in units of 1m for dimensional reasons, and $\omega=h_c/h$. Recall that $\ell_i$ is the plate length in the $i=L,T$ direction.
The ratio of the half-wave lengths $n_T$ and $n_L$ becomes
\be
\dfrac{n_T}{n_L}=  \sqrt[4]{\dfrac{E_L(T^*)}{E_T(T^*)}}\,,
\ee
which could be used to validate the proposed model, once the high temperature material properties are available with sufficient accuracy.
%
%
%
%

We now continue our investigation by addressing the phenomenon from a strictly mechanical viewpoint.
We want to evaluate the combined stress state at the nodes of the buckling modes and prove that an \textit{extensional principal stress} at such nodes always exists. The stress state is here determined as a combination of compressive membrane stress (thermal stress) and a non-vanishing vertical shear stress\footnote{The following argumentation is based on thin plate theory, which is extensively discussed in \cite{ventsel2001thin}.}.
 
Consider a differential beam strip element $\epsilon dn$ illustrated in Fig.~\ref{fig:Extensional_stress}
and a point $P$ on the node-lines. The planes (or sections) normal to $n$, $t$ and $z$ are 
denoted by $\Pi_n$, $\Pi_t$ and $\Pi_z$, respectively. The buckling mode $w$ with origin at $P$ is obtained as a solution of the eigenproblem Eq.(\ref{coupledproblem}). Its local form is
\be
w(n)= w_0 \sin\left( \dfrac{2\pi n}{\lambda}\right),
\ee
where $\lambda$ is the wavelength and $w_0 \ne 0$ is the maximum off-plane displacement \cite{timoshenko2012theory}. In the following, $w_{tt}\equiv\partial^2w/\partial t^2$ and so on.

At the node-lines, we have zero displacement, $w(P)=0$, and there is no pressure from the substrate 
($r(P)= kw(P) = 0$) and no normal stress component $\sigma_{zz} = 0$ on the plane $\Pi_z$. 
In addition, both bending curvatures vanish at node $P$
\bea
&&\kappa_n = -w_{nn}(P)=0\,,\\
&&\kappa_t = -w_{tt}(P)=0\,,
\eea
which implies $M_n=M_t=0$ for both bending moments and $\sigma_{nn}=\sigma_{tt}=0$ for the respective bending normal stress components for planes $\Pi_n$ and $\Pi_t$.
Accordingly, only the compressive in-plane normal thermal stresses are non-zero,
\be
\sigma^N_{nn} 
\equiv -\sigma_N  = - E \alpha_R \Delta T < 0\,,
\ee
for the plane $\Pi_n$ with Young modulus $E$, and 
$
\sigma^N_{tt} < 0
$
for the plane $\Pi_t$.

We now compute the shear forces, leading to the shear stress in $\Pi_n$. The shear force $Q_n$ is non-zero along $t$, since 
\bea
&&w_{tt} \equiv 0 \Rightarrow (w_{tt})_{,n} = \partial_n (w_{tt})=0,
\\ 
&&(w_{nn})_{,n} = -w_0 \left(\dfrac{2\pi}{\lambda}\right)^3 \ne 0.
\eea
Therefore the following holds for the shear force 
\be
Q_n = -(D_n w_{nn})_{,n} \ne 0\,,
\ee
with bending rigidity $D_n$ along the $n$-direction. Moreover, $Q_n$ is of maximum amplitude since $\cos (n \pi/\lambda) = 1$ at $P$.

The corresponding shear stress is $\tau_{nz} \propto Q_n$, thus its average $|\tau(P)| \ne 0$ is maximal as well, while the average shear stress is integrated over the area $A_c$ as
\be
\tau \equiv \int_{A_c}\tau_{nz} dA_c/A_c = Q_n/A_c\,.
\ee
%
On the plane $\Pi_t$, the shear force is instead
\be
Q_t = -(H w_{nn} + D_t w_{tt})_{,t} = 0,
\ee
since  $w_{tt} \equiv 0$ and $(w_{nn})_{,t}$ vanishes. In other words, the shear stress components vanish: $\tau_{tz}= \tau_{zt}=0$.

%
On the plane $\Pi_n$, the twist rate is
\be
\partial_n (w_{t}) = w_{tn} = 0,
\ee
meaning that also the twist moment $M_{nt}=0$ and the shear stress $\tau_{nt}=0$. By symmetry of the stress tensor, $\tau_{tn}=0$ and $M_{tn}=0$ on the plane $\Pi_t$. This implies that $\Pi_t$ is a principal stress plane, because only the normal compressive membrane stress $\sigma_T < 0$ is non-zero \cite{ventsel2001thin,timoshenko2012theory}.
Now, since
\be
\partial_{n}w_{nn} = -w_0\left( \dfrac{2\pi n}{\lambda}\right)^3 \ne 0,
\ee
the shear force is non-zero as well,
\be
Q_n \propto-\partial_{n}w_{nn}\ne 0.
\ee
Moreover, its amplitude is maximal, since $\cos (n \pi/\lambda) = 1$ at $P$.
The average shear stress is now
\be
\tau = \dfrac{Q_n}{A_c}\ne 0\,.
\ee
At point $P \in \Pi_n$, a combined stress-state
 \be\label{stressstate}
 {\boldsymbol \sigma} = \begin{bmatrix}
 	-\sigma_{N} & \tau  \\
 	\tau       &  0    \\
 \end{bmatrix}
 \ee
thus exists, with non-zero shear stress $\tau$ of maximal magnitude, and a compressive membrane stress  $\sigma_N < 0$.

Since one principal stress plane is known, it is sufficient to investigate the remaining two, which are mutually perpendicular to $\Pi_t$.
The principal stresses $\sigma_1$, $\sigma_{2}$ are the two eigenvalues  of the stress tensor $\boldsymbol \sigma$,
\be
 \sigma_{1,2} = \frac{\sigma_N}{2} \left(-1\pm \sqrt{1 + 4 \dfrac{\tau^2}{\sigma_N^2}}\,\right)\,,
\ee
meaning that the major extensional principal stress $\sigma_1 > 0$ since $\tau \ne 0$.

Therefore we have proven that a state of extensional principal stress always exists perpendicular to each point of the node-line (Fig.~\ref{fig:Extensional_stress}). The cracks will appear at these locations, as was shown by the numerical solution in Fig.~\ref{fig:orthotropicvonmises}c. 

Physically, the thermomechanical instability phenomenon is a consequence of two simultaneous changes in the thermophysical properties of wood: softening and an increasing thermal expansion coefficient. Wood is an elasto- viscoplastic natural composite with hierarchical cellular structure, consisting of cellulose microfibrils embedded in a lignin- hemicellulose matrix. 
Cellulose is crystalline, while lignin and hemicellulose are amorphous natural polymers with glass transition temperatures $T_g \approx 180 - 200^\circ{\mathrm C}$ for dry wood and $T_g \approx 100 ^\circ{\mathrm C}$ at 10 \% moisture content \cite{Salmen_1979,Salmen_1984_bis,Bazant_1985,Salmen_1984,Antoniow2012}.

\begin{figure}[t]
	\centering
	\includegraphics[width=0.4\textwidth]{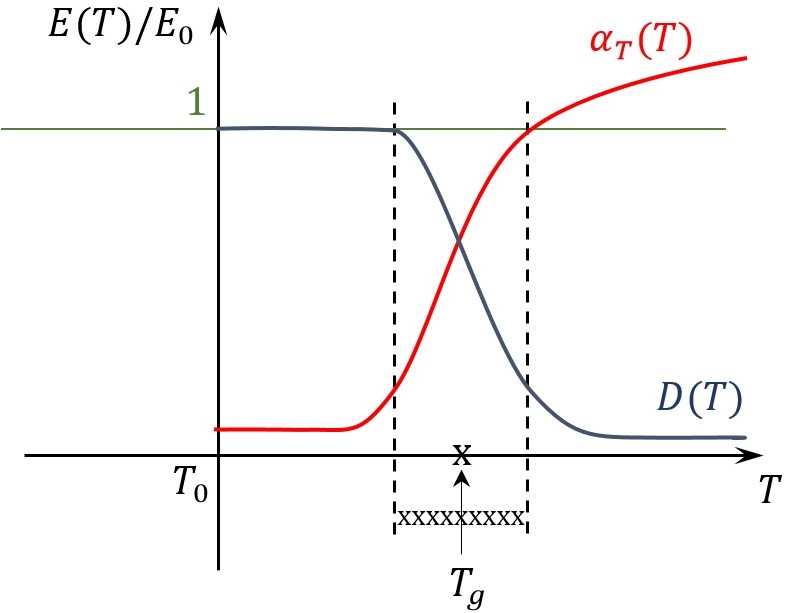}
	\caption{Thermal dependence of elasticity modulus $D(T)$ and thermal expansion coefficient $\alpha_i(T)$ for wood.}
	\label{fig:Elasticityvsthermalexp}
\end{figure}

Below the glass transition temperature $T_g$, wood is hard (glassy-state).
Around the glass transition temperature $T_g$, the thermal expansion coefficient $\alpha_i$ increases dramatically, while the elastic coefficients simultaneously decrease (Fig.~\ref{fig:Elasticityvsthermalexp}). Above $T_g$, wood enters the rubbery state and softens critically. In fact, the thermal expansion is often used to identify the glass transition temperature region in experiments \cite{Antoniow2012}. The increment in $\alpha_i$ induces substantial thermal stresses due to the restrained thermal elongation from the cold layer.

The softening takes place below the thermal degradation (or pyrolysis) temperature $T_p\approx 300 ^\circ{\mathrm C}$. 
For the Medium Density fibreboard (MDF = wood composite: $>80\%$ wood fibres, 10\% resin and wax adhesive, 10\% water), the softening mechanism occurs for temperatures within $\approx 75 - 125 ^\circ{\mathrm C}$ \cite{Jian_et_al_1984}. Due to the high wood fibre content, we expect that the physical explanation for wood also applies to MDF.

\section{Conclusions}\label{sec:conclusions}
 
Until now it was believed that the characteristic crack pattern on the surface of charred wood and other cellulosic materials are created by physicochemical processes which occur during and beyond pyrolysis. In this paper we argued that the observed formation of the global macro-crack patterns below the pyrolysis temperature is due to the thermomechanical effects that induce wrinkling or buckling of the heated surface.
The physical explanation follows from the thermoplastic properties of the heated wood surface: as the hot layer softens and elongates significantly when approaching the glass transition temperature, the cold substrate layer restrains the thermal elongation and induces substantial thermal stresses. 

Despite the lack of accurate mechanical material properties, the 3D thermomechanical buckling model could predict the locations of the buckling node-lines consistently with the experimentally observed major crack-lines. The model was able to reproduce the qualitative difference between the crack patterns of the orthotropic (fir) and isotropic (MDF) materials. By assuming different values for the relative softening at high temperature, we were able to even predict the non-trivial topologies observed experimentally for MDF after the onset of pyrolysis.

Through a simplified (2D) model, we showed indeed that the cracking will occur along the node-lines of the buckling modes because there always exists a major extensional stress, both along and orthogonally to such node-lines. According to our model, this is a necessary condition for buckling. Using the simplified model, we were able to develop two predictive formulas: one for the inter-node distance (wave length) of the buckling mode, and one for the buckling load, which returns the value of the driving force (thermal stress) above which the plane surface wrinkles and buckles.
These results can be used to find new ways to improve the fire safety of wood-based materials by preventing, reducing or delaying the crack formation.
  
Although these promising results were obtained, several sources of uncertainty can be identified. First of all, the mechanical properties of the sample materials at sufficiently high temperatures are not yet available, and a coupled thermomechanical analysis with thorough validation would not be justified. Moreover, competing theories of the cracking phenomenon, mainly based on the shrinkage assumption, can still be proposed and supported by experimental observations from different heating conditions. The processes of crack nucleation, growth and propagation were not investigated here, nor did we include burning or actual charring process which would dramatically affect the tensional strength of the surface layer. As always, the real char cracking phenomenon can indeed be a combination of several competing mechanisms, and much more complex than our models.

\section*{Aknowledgements}

 The authors would like to thank Reijo Kouhia for comments and suggestions. This work was supported by the National Natural Science Foundation of China (NSFC) under Grant No. 51406192 and by the Academy of Finland under Grant No. 297030.

\appendix

\section{Analytical solutions of buckling modes}
\label{sec:analytical-solution-for-buckling-modes-of-a-thin-wood-layer}


\subsection{Buckling equation}

For a dry material below its pyrolysis temperature, where both evaporation and degradation -induced shrinkage can be ignored, the total potential energy $\bm{\Pi}$ of the system in Fig.~\ref{fig:Thinplate} can be written as
\be\label{totalenergyintegral}
\bm{\Pi}=\mathbb{U}-\mathbb{W}\,,
\ee
where the internal energy of the hot layer of area $A$ is
\bea\label{U}
\mathbb{U}&=&\frac{1}{2}\int_A
\left(
D_xw^2_{xx}+D_yw^2_{yy}+
2D_{xy}w_{xx}w_{yy}
\right.\nonumber\\
&&\left.
+4D_Sw^2_{xy}
\right)dA
+\frac{1}{2}\int_A k(x,y)w(x,y)^2 dA
\,,\nonumber\\
\eea
and the work of external (membrane- and pressure-) forces 
\be\label{W}
\mathbb{W}=\frac{1}{2}\int_A
\left( N_{xx}^0 w^2_{xx}+N_{yy}^0 w^2_{yy}\right)
dA
+\int_Ap^0(x,y)w(x,y)dA
\,.
\ee
Here $w_{\alpha\beta}(x,y)=\partial^2w/\partial\alpha\partial\beta$ is the $\alpha,\beta$ element of the curvature matrix, $D_S$ is the shear modulus and the apex 0 refers to values immediately before the buckling takes place.
In Eq.(\ref{U}) we recognize the separate contributions from the elastic thin plate and the substrate, modelled as a set of harmonic oscillators with spring constant $k$ given by Eq.(\ref{springcoeff}).

The stability criterion for this configuration is given by the Trefftz condition (see e.g. \cite{ventsel2001thin} and references quoted therein). Imposing the differential variation of the total energy of the system to be identically zero,
\be\label{Trefftz}
\delta\bm{\Pi}=\delta\mathbb{U}-\delta\mathbb{W}=0\,,
\ee
and expanding this condition through (\ref{U}) and (\ref{W}), we get the differential equations for the buckling problem (or buckling equations),
\bea\label{coupledproblemexpanded}
&& D_{xx}w_{xxxx}+2(D_{xy}+2D_S)w_{xxyy}+D_{yy}w_{yyyy}
\nonumber\\
&&+(N_{xx}^0w_{xx}+N_{yy}^0w_{yy})
\nonumber\\
&& \equiv 
D_{xx}w_{xxxx}+H w_{xxyy}+D_{yy}w_{yyyy}
\nonumber\\
&&
+(N_{xx}^0w_{xx}+N_{yy}^0w_{yy})
=p(x,y)-kw(x,y),\nonumber\\
\eea
with the deformation field $w(x,y)$ as solution. The above (\ref{coupledproblemexpanded}) is the most general differential equation for this mechanism, for both orthotropic and isotropic materials and including torsion and pressure forces acting on the thin plate.

\subsection{Buckling modes with no torsion}\label{Buckling modes with no torsion}

In this section we start from Eq.(\ref{coupledproblemexpanded}) to derive the buckling equations (\ref{coupledproblem}) for a torsion-free orthotropic plate.
%
%
%
Focusing on the $x$ axis, and assuming negligible torsion (i.e., set $D_{xy}$ and $w_{xy}$ both $\sim 0$), we rewrite Eq.(\ref{coupledproblemexpanded}) as
\be
(D_{xx}w^{\prime\prime})^{\prime\prime}+N_{xx}^0w^{\prime\prime}+kw=p\,.
\ee
The buckling here is not caused by pressure forces, but only by the \textit{thermally-driven} increment of the membrane forces $N_{xx}^0$. Since $p=0$, we finally obtain
\be\label{bucklingproblem}
w^{(4)}+\lambda^2 w^{\prime\prime}+\beta^4 w =0\,,
\ee
where $\lambda^2\equiv \mathbf{P}/D_{xx}$, $\mathbf{P}\equiv N_{xx}^0$ and $ \beta^4\equiv k/D_{xx}$.

By using now the Rayleigh-Ritz method (see for instance \cite{Leissa2005961,Trefethen:1997:NLA,trove.nla.gov.au/work/6379829}), we expand the displacement field in Fourier series
\be
w(x)=\sum_{n=1}^\infty a_n \sin{\left( \frac{n\pi}{\ell_i} x\right)}\,,
\ee
and substitute in Eq.(\ref{bucklingproblem}). $\ell_i$ is the plate length in the $i=L,T$ direction. Solving for $\lambda$ gives the criticality condition
\be
\lambda_{cr}^2=\left( \frac{n\pi}{\ell_i} \right)^2+\left( \frac{\ell_i}{n\pi} \right)^2\beta^4
\equiv \mathbf{\Lambda}(n)\,.
\ee
$\mathbf{\Lambda}(n)=\mathbf{P}/D_{xx}$ is a control parameter, which when minimized gives the number of nodes that correspond to the first cracks. 
We easily find
\be
n_i=\pm\beta\left(\frac{\ell_i}{\pi}\right)\,, \qquad n_i\in \mathbb{N}\,,
\ee
which can be recast in terms of the material properties,
\be\label{numberofnodes}
n_{L,T}(\omega)=\frac{\sqrt[4]{12}}{\pi}
\sqrt[4]{\frac{E_R(T_0)}{E_{L,T}(T_*)}}\left(\frac{\ell_{L,T}}{h}\right)
\frac{1}{\omega^{3/4}}\sqrt[4]{\frac{h}{1{\mathrm m}}}\,,
\ee
by expanding $\beta$ as
\be
\beta= \sqrt[4]{ \left( \frac{k}{D_{xx}}\right)} =
\frac{\sqrt[4]{k}}{\sqrt[4]{E_{xx}}h_c^{3/4}} \sqrt{6}
=\frac{\sqrt[4]{k}}{\sqrt[4]{E_{L,T}}h_c^{3/4}} \sqrt{6}\,,
\ee
and recalling the definition of shear modulus $G=E/2(1+\nu)\,[N/m^2]$.
%
%
The according critical thermal stress is 
\bea\label{criticaldeformations}
&&(N_{xx})_{cr}=\frac{\mathbf{P}}{D_{xx}}=\left(\frac{\pi}{\ell_{L,T}} \right)^2
\times
\nonumber\\
&&
\times
\Big[ n_{L,T}^2+\frac{k}{D_{xx}}
\left(\frac{\ell_{L,T}}{\pi} \right)^4\frac{1}{n_{L,T}^2}\Big] \equiv 
\mu_{cr}\left(\frac{\pi}{\ell_{L,T}} \right)^2\,,\nonumber\\
\eea
where $(N_{xx})_{cr} = E_{xx}\alpha_{xx} (T^*-T_0) A_c$ from the constitutive relation between stress and strain.

%
%
%

\subsection{Buckling modes, torsion-coupled}\label{Buckling modes, torsion-coupled}

If we include the effect of torsion on the thin hot layer, all the terms in the buckling equation (\ref{coupledproblemexpanded}) must be considered.
Substituting the displacement fields
\be
w(x,y)=\sum_{m=1}^\infty\sum_{n=1}^\infty a_{m,n}\sin{\left( \frac{m\pi x}{\ell_x}\right) }
\sin{\left( \frac{n\pi y}{\ell_y}\right) }\,,
\ee
and setting again $p=0$, we obtain the following critical thermal stresses:
\be\label{deformationstorsion}
(N_{xx}^0)_{cr}=\frac{D_{xx}\alpha_n^4+D_{xx}\beta_m^4+H\alpha_n^2\beta_m^2+D_{xx}\beta^4}{\alpha_n^2+\beta_m^2}\,.
\ee
The above reduces to the decoupled case Eq.(\ref{criticaldeformations}) if $\beta_m=0$. Here $\alpha_n\equiv n\pi/\ell_x$ and $\beta_m\equiv m\pi/\ell_y$.
Now, define $D_0\equiv D_{xx}$ and
\be
\eta_{TL}=\frac{D_{yy}}{D_{xx}}\,,\qquad \eta_H=\frac{H}{D_{xx}}\lesssim0.24\,,
\ee
so the critical thermal stresses (\ref{deformationstorsion}) can be rescaled as
\be
\frac{(N_{xx}^0)_{cr}}{D_0}=\frac{\alpha_n^4+\beta_m^4+\eta_H\alpha_n^2\beta_m^2+\beta^4}{\alpha_n^2+\lambda\beta_m^2}\equiv f(m,n)\,,
\ee
(remember that $\lambda^2\equiv \mathbf{P}/D_{xx}$). Minimizing this critical buckling load $f(m,n)$ with respect to $m$ and $n$ gives the location of the cracking patterns observed in the experiments. Keeping $m$ fixed, some tedious algebra gives
\bea\label{nalphatorsion}
&n(m,\omega)=\dfrac{m}{\sqrt{2}}\left( \dfrac{\ell_x}{\ell_y}\right) 
\left[
-1+\left[
\half+48m^4\left(\dfrac{\ell_y}{\pi}\right)^4
\times\nonumber
\right.\right.\\
&\left.\left.
\times
\dfrac{E_R(T_0)}{E_L(T_0)}\dfrac{E_L(T_0)}{E_L(T^*)}
\dfrac{1}{\omega^3h^3}+4\dfrac{E_T(T^*)}{E_L(T^*)}\right]^{1/2}
\right]^{1/2}
\nonumber\\
&
\approx
\dfrac{m}{\sqrt{2}}\sqrt{
	-1+\sqrt{0.7+0.00123m^4\dfrac{R}{\omega^3}}
}\,,
\eea
where we expanded all the coefficients in terms of the material parameters and recalling that $\lambda=\sqrt{ \mathbf{P}/D_{xx}}\approx1/2$. Also, $\ell_x=\ell_y=100$mm and $h=25$mm. The numerical approximation holds by virtue of the Young moduli values for typical woods, namely $E_R(T_0)/E_L(T_0)\sim 0.1$ and 
$E_T(T^*)/E_L(T^*) \sim 0.05$ on the average \cite{Woodhandbook}.
\begin{figure}[t]
	\centering
	\includegraphics[width=0.45\textwidth]{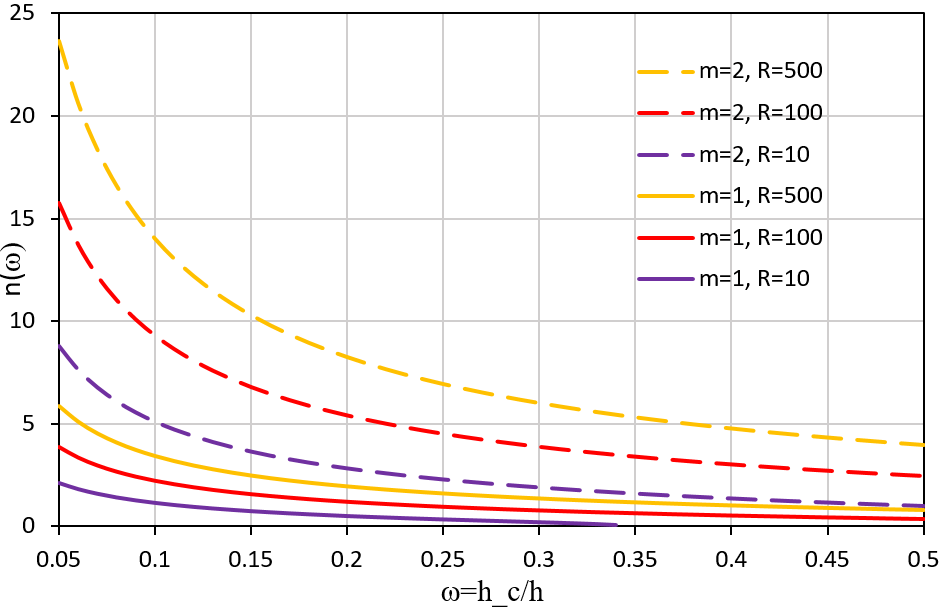}	
	\caption{Number of nodes in function of the relative thickness $\omega$ of the thin hot layer, for $m=1$ (solid) and $m=2$ (dashed).
	}
	\label{fig:N_fun_torsion}
\end{figure}
Reality of Eq.(\ref{nalphatorsion}) depends on the term $R/\omega^3$, with
\be
R\equiv \frac{E_L(T_0)}{E_L(T^*)}\,,
\ee
the ratio of the Young moduli along the longitudinal direction for $T_0$, the temperature of the cold layer, and $T^*$, that of the hot layer. The reality condition is satisfied identically $\forall\omega\in(0,1]$ if $R>244$.
Otherwise, we find the following constraints:
\bea
&&
R=5 \Rightarrow \omega<0.27\,,\\
&&
R=10 \Rightarrow \omega<0.34\,,\\
&&
R=100 \Rightarrow \omega<0.74\,.
\eea
We plotted Eq.(\ref{nalphatorsion}) in Fig.~\ref{fig:N_fun_torsion} for $m=1$ and $m=2$, with $R$=10, 100, 500. In terms of the $R$-parameter and material properties, the number of nodes for the decoupled case (\ref{numberofnodes}) can also be rewritten as
\be
n_{L,T}(\omega)=0.5295\left(\dfrac{R}{\omega^3} \right)^{1/4}\,.
\ee
It can be shown that in Fig.\ref{fig:N_fun_torsion}, for a fixed $R$ the curves in the two cases overlap for $m=2$:  $n(m=2,\omega)=n_{L,T}(\omega)$.

\section*{References}

\bibliography{thermo-mechanical}

\end{document}